# High-speed optical correlation-domain reflectometry without using acousto-optic modulator

Makoto Shizuka, Shumpei Shimada, Neisei Hayashi, Yosuke Mizuno,* and Kentaro Nakamura

*Precision and Intelligence Laboratory, Tokyo Institute of Technology, Yokohama 226-8503, Japan*
*ymizuno@sonic.pi.titech.ac.jp*

## Abstract

To achieve a distributed reflectivity measurement along an optical fiber, we develop a simplified cost-effective configuration of optical correlation- (or coherence-) domain reflectometry based on a synthesized optical coherence function by sinusoidal modulation. By excluding conventional optical heterodyne detection (practically, without using an acousto-optic modulator) and by exploiting the foot of the Fresnel reflection spectrum, the electrical bandwidth required for signal processing is lowered down to several megahertz. We evaluate the basic system performance and demonstrate its high-speed operation (10 ms for one scan) by tracking a moving reflection point in real time.

## 1. Introduction

Optical reflectometry is a useful technology for health monitoring of optical components, modules, and fiber networks, and also serves as a fundamental technique for a variety of multiplexed and distributed sensing systems [1–5] and optical coherence tomography [6–8]. To date, to detect bad connections/splices and other reflection points along fibers under test (FUTs), following two kinds of Fresnel-based fiber-optic reflectometers have been developed and extensively studied: optical time-domain reflectometry (OTDR) [9–13] and optical frequency-domain reflectometry (OFDR) [14–18]. However, in general, OTDR suffers from a relatively low spatial resolution, a low sampling rate (or a long measurement time), while OFDR suffers from phase fluctuations caused by environmental disturbance. One promising method for overcoming these drawbacks is so-called optical correlation (or coherence)-domain reflectometry (OCDR) [19–28] exploiting synthesized optical coherence function (SOCF) [26], which operates based on the correlation control of propagating lightwaves by modulating the frequency of the laser output. Two methods for the frequency modulation have been implemented so far: sinusoidal [21–23] and stepwise modulations [24–26]. The latter includes the modulation using an optical frequency comb [27, 28], which can enhance the measurement stability. Of these two methods, sinusoidal modulation is the more suitable for cost-efficient implementation.

In conventional OCDR-SOCF systems [21–28], the Fresnel spectra to be measured are shifted by several tens of megahertz by optical heterodyne detection using acousto-optic modulators (AOMs). This is because otherwise low-frequency noise of the electrical devices overlaps the Fresnel spectra. If the system can be implemented without the use of optical frequency shift, electrical signal processing can be performed in the frequency range near direct current (DC), leading to cost reduction of the relevant devices and thus of the whole system.

In this work, by excluding optical heterodyne detection, we newly develop a simple configuration of OCDR-SOCF based on sinusoidal modulation. By exploiting the foot of the Fresnel reflection spectrum, the system is simplified with reduced cost. Its high-speed operation is also demonstrated.

## 2. Principle

The experimental setup of standard OCDR-SOCF containing an AOM [21–23] is depicted in Fig. 1(a). The laser output is divided with a coupler; one is injected into an FUT and the other is used as reference light. The reflected light from the FUT is mixed with the reference light, the frequency of which is downshifted by several tens of megahertz using the AOM for optical heterodyne detection. The signal is then converted into an electrical signal with a photo detector (PD), and is monitored using an electrical spectrum analyzer (ESA). Owing to the optical heterodyne detection, the influence of the low-frequency noise of the PD and ESA can be mitigated. To achieve high-speed operation, the spectral peak power is continuously recorded using the analog output terminal of the ESA [23].

To spatially resolve the measurement positions, we sinusoidally modulate the frequency of the laser output to form a so-called "correlation peak" in the FUT, with which the reflected light generated at a specific position is selectively observed [26]. By sweeping the modulation frequency, the correlation peak is scanned along the FUT, and thus the reflectivity can be measured in a distributed manner. Sinusoidal frequency modulation generates multiple correlation peaks periodically, the interval of which determines the measurement range. According to detailed calculations, the spatial resolution $\Delta z$ (i.e. the 3-dB linewidth of the correlation peak) and measurement range $D$ are theoretically given by [29]

$$\Delta z \cong \frac{0.76c}{\pi n \Delta f}, \qquad (1)$$

$$D = \frac{c}{2nf_m}, \qquad (2)$$

respectively, where $c$ is the light velocity in vacuum, $n$ is the refractive index of the fiber core, $\Delta f$ is the modulation amplitude, and $f_m$ is the modulation frequency.

In contrast, the experimental setup of a new OCDR-SOCF system without optical heterodyne detection is shown in Fig. 1(b), which is basically the same as Fig. 1(a) except that no AOM is contained. If the Fresnel reflection spectrum has a delta-function-like shape, this configuration does not work properly. However, as the Fresnel spectrum in reality has some non-negligible bandwidth (determined by the laser bandwidth) [30], the foot of the Fresnel spectrum can avoid the overlap of the low-frequency noise of the PD and ESA even without the use of optical frequency shift. In this configuration, the frequency at which to measure the optical power ranges from several hundreds of kilohertz to several megahertz. The theoretical expressions for the spatial resolution and measurement range are the same as those in the aforementioned standard OCDR-SOCF.

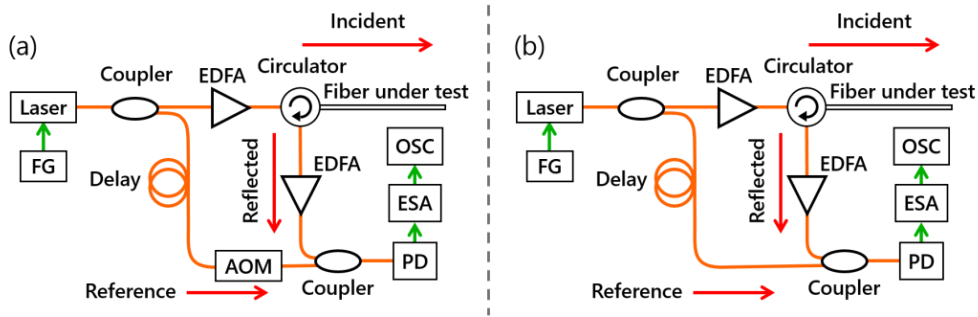

Fig. 1. Experimental setup of optical correlation-domain reflectometry (OCDR). (a) A conventional system and (b) a new system without an acousto-optic modulator (AOM). EDFA, erbium-doped fiber amplifier; ESA, electrical spectrum analyzer; FG, function generator; OSC, oscilloscope; PD, photo detector.

### 3. Experiments

In the experiments below, we employed a laser diode at 1550 nm with a 3-dB bandwidth of ~1 MHz, and its output was amplified to ~19 dBm and then injected into FUTs. The modulation amplitude $\Delta f$ was fixed at 0.75 GHz (to avoid the damage to the laser), resulting in the theoretical spatial resolution of approximately 66 mm (see Eq. (1)). Note that a spatial resolution of higher than 100 μm is theoretically achievable by employing a high-speed broadband tunable laser, such as a superstructure-grating distributed Bragg reflector laser.

First, to confirm the basic operation of the OCDR without optical heterodyne detection, distributed reflectivity measurements were performed using a simply structured FUT depicted in Fig. 2. A 1.0-m-long pigtail of a circulator composed of a silica single-mode fiber (SMF) was connected to 1.0- and 3.0-m-long silica SMFs sequentially using angled physical contact (APC) connectors; the end of the 3.0-m-long SMF was kept open also with an APC connector. The reflectivity distribution along the FUT at a specific frequency $f_z$ was derived from the ESA and observed using an oscilloscope (OSC). The resolution bandwidth (RBW) and the video bandwidth (VBW) of the ESA were set to 300 kHz and 1 kHz, respectively. The modulation frequency $f_m$ was swept from 5.01 MHz to 5.16 MHz with a repetition rate of 10 Hz, corresponding to the measurement range of approximately 20 m according to Eq. (2).

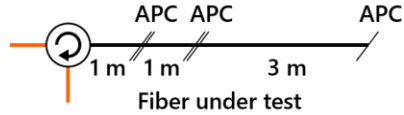

Fig. 2. Structure of the fiber under test for basic characterization.

The measured electrical spectrum in the range from DC to 12.0 MHz (when $f_m = 5.1$ MHz) is shown in Fig. 3. Besides the low-frequency noise around DC, two peaks were observed at 5.1 MHz and 10.2 MHz, which corresponds to $f_m$ and its second harmonic component, respectively. The reflectivity distributions obtained when $f_z = 1.2$, 2.0, and 5.0 MHz are then shown in Fig. 4(a), (b), and (c), respectively. The relative position $d$ was defined to be 0 at the circulator. In Figs. 4(a) and (b), three clear peaks with higher than −52-dB reflectivities were observed, the locations of which well corresponded to the APC connectors (including the open end). The relatively small peak (at $d = 0$) corresponds to the reflection at the circulator (approximately −52.5 dB). In contrast, in Fig. 4(c), no peaks were observed because of the overlap of the noise component induced by the modulation frequency $f_m$. Next, the signal-to-noise ratio (SNR), defined as the difference between the maximal peak power and the noise floor level, was plotted as a function of $f_z$ (Fig. 5). When no peaks correspondincomg to the APC connectors were observed, the SNR was defined to be 0. The SNR was deteriorated at $f_z$ < ~1 MHz in this condition (dependent on the RBW and VBW) because of the low-frequency noise of the ESA and PD. The SNR was also reduced at $f_z$ of around 5 MHz because of the noise caused by the laser modulation. The maximal SNR was obtained at $f_z$ ~2 MHz.

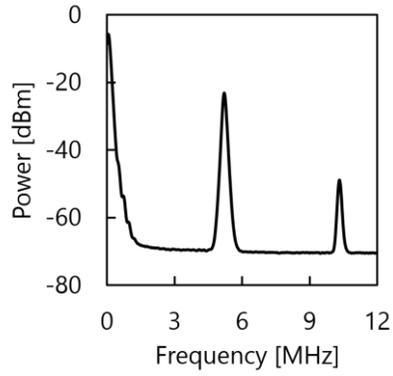

Fig. 3. Electrical spectrum measured when $f_\mathrm{m}$ was 5.1 MHz.

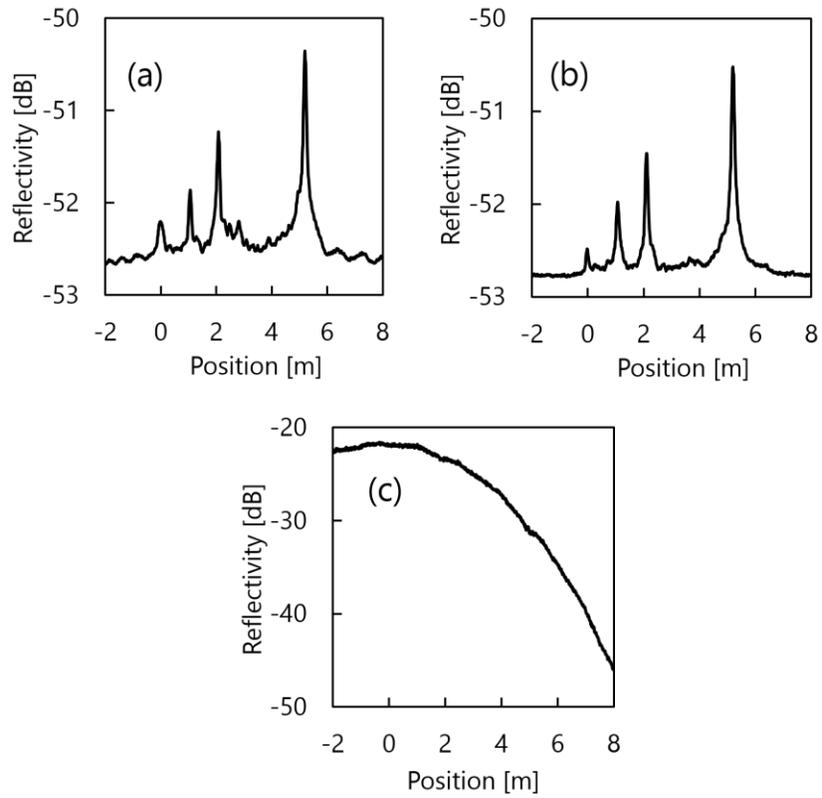

Fig. 4. Reflectivity distributions measured when $f_z$ was (a) 1.2 MHz, (b) 2.0 MHz, and (c) 5.0 MHz.

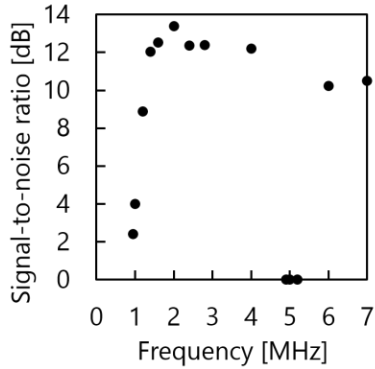

Fig. 5. Signal-to-noise ratio (SNR) plotted as a function of $f_z$.

The relation between the peak power obtained by this method and the actual reflection power (measured using a power meter) was also investigated (Fig. 6). According as the reflection power increased, the peak power also increased monotonically. The dependence was not linear probably because of the nonlinear characteristics of the PD. Nevertheless, by using this one-to-one correspondence, the actual reflection power can be inferred from the measured peak power.

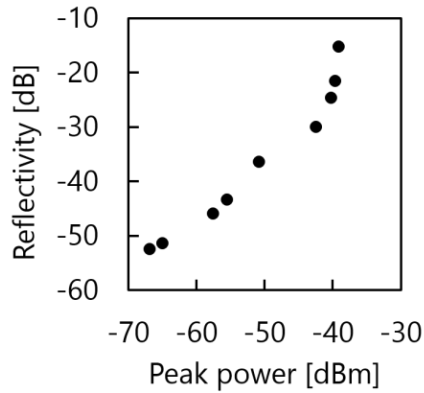

Fig. 6. Actual reflection power vs. peak power obtained by this method.

Subsequently, we performed the same measurement using another FUT with more complicated structure (Fig 7). A 1.0-m-long pigtail of the circulator was connected to 1.0-, 1.0-, 3.0-, and 1.0-m-long silica SMFs sequentially using APC connectors; the distal end of the FUT was kept open also with an APC connector. The modulation frequency $f_m$ was swept from 5.01 MHz to 5.20 MHz with a repetition rate of 10 Hz ($D$ = ~20 m). Fig. 8 shows the reflectivity distribution ($f_z$ = 2.0 MHz). Five clear peaks were observed, the locations of which well corresponded to the APC connectors (including the open end). A relatively small peak corresponding to the reflection at the circulator was also observed.

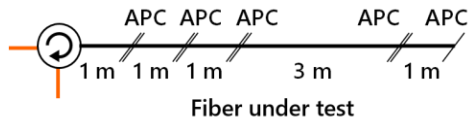

Fig. 7. Structure of the fiber under test for the demonstration of numerous reflection point detection.

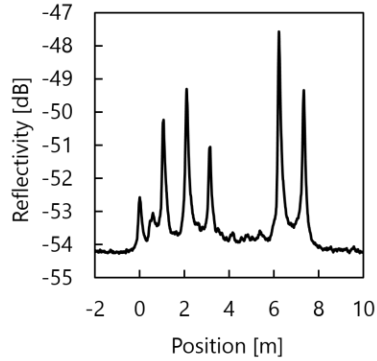

Fig. 8. Reflectivity distribution measured when $f_s$ was 2.0 MHz.

Finally, high-speed operation of this system was verified by tracking a moving reflection point. The structure of the FUT is shown in Fig. 9. Part of a 7.0-m-long silica SMF (with a 0.3-mm-thick jacket) was wound on a screw with an outer diameter of 6.4 mm (fixed on a stage), and the stage was moved along the SMF for ~3.0 m at a constant speed of 0.2 m/s. Thus, a moving reflection point can be implemented, though the reflectivity is not sufficiently stable. The reflectivity distributions measured when time $t = 0$, 3.0, 6.0, and 9.0 s are shown in Fig. 10(a) ($t$ was defined to be 0 when the screw started to move). In addition to the fixed peak at the APC connector, another peak was clearly observed at different positions. The peak power of ~53.5 dB agreed well with the actual value (~53 dB). The peak power fluctuated within approximately +/−1 dB, which is partially attributed to the actual reflectivity fluctuations. The temporal variation of the measured reflectivity distribution is then shown in Fig. 10(b). We clearly recognize the linearly moving reflection point; the moving speed was calculated from the measured data to be 0.19 m/s, which is in good agreement with the actual value.

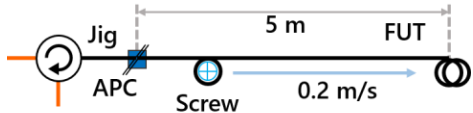

Fig. 9. Structure of the fiber under test for the demonstration of high-speed operation.

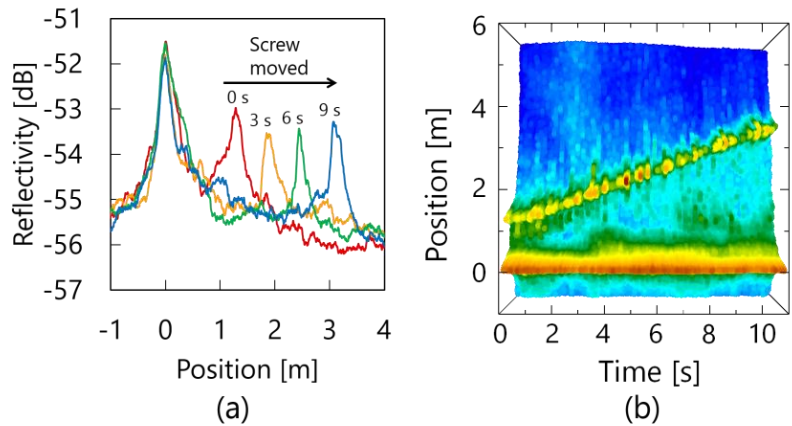

(a)                                    (b)

Fig. 10. (a) Reflectivity distributions measured when time $t = 0$, 3.0, 6.0, and 9.0 s. (b) Temporal variation of the measured reflectivity distribution.

## 4. Conclusion

In this work, a simplified cost-effective configuration of OCDR-SOCF based on sinusoidal modulation was developed by excluding conventional optical heterodyne detection, and its basic performance was evaluated. By the use of the foot of the Fresnel reflection spectrum, the electrical bandwidth required for signal processing was lowered down to several megahertz. A high-speed reflectivity measurement (10 ms for acquiring one distribution) was also demonstrated by tracking a moving reflection point. Further cost reduction will be possible by replacing the ESA with an electrical circuit with the equivalent function (narrow band-pass filtering, etc). We believe that this configuration will boost the convenience of the OCDR-SOCF technology for practical use in the future.

## Acknowledgments

This work was supported by JSPS KAKENHI Grant Numbers 25709032, 26630180, and 25007652, and by research grants from the Iwatani Naoji Foundation, the SCAT Foundation, and the Konica Minolta Science and Technology Foundation.

## References

1. A. H. Hartog, A. P. Leach, and M. P. Gold, "Distributed temperature sensing in solid-core fibres," Electron. Lett. **21**, 1061–1062 (1985).
2. P. K. C. Chan, W. Jin, and J. M. Gong, "Multiplexing of fiber Bragg grating sensors using a FMCW technique," IEEE Photonics Technol. Lett. **11**, 1470–1472 (1999).
3. Y. Mizuno, W. Zou, Z. He, and K. Hotate, "Proposal of Brillouin optical correlation-domain reflectometry (BOCDR)," Opt. Express **16**, 12148 9–12153 (2008).
4. L. Palmieri and A. Galtarossa, "Distributed polarization-sensitive reflectometry in nonreciprocal single-mode optical fibers," J. Lightwave Technol. **29**, 3178–3184 (2011).
5. W. Zou, S. Yang, X. Long, and J. Chen, "Optical pulse compression reflectometry: proposal and proof-of-concept experiment," Opt. Express **23**, 512–522 (2015).
6. D. Huang, E. A. Swanson, C. P. Lin, J. S. Schuman, W. G. Stinson, W. Chang, M. R. Hee, T. Flotte, K. Gregory, C. A. Puliafito, "Optical coherence tomography," Science **254**, 1178–1181 (1991).
7. M. A. Choma, M. V. Sarunic, C. Yang, and J. A. Izatt, "Sensitivity advantage of swept source and Fourier domain optical coherence tomography," Opt. Express **11**, 2183–2189 (2003).
8. Y. Kato, Y. Wada, Y. Mizuno, and K. Nakamura, "Measurement of elastic wave propagation velocity near tissue surface through optical coherence tomography and laser Doppler velocimetry," Jpn. J. Appl. Phys. **53**, 07KF05 (2014).
9. M. K. Barnoski and S. M. Jensen, "Fiber waveguides: a novel technique for investigating attenuation characteristics," Appl. Opt. **15**, 2112–2115 (1976).
10. G. P. Lees, H. H. Kee, and T. P. Newson, "OTDR system using Raman amplification of a 1.65 µm probe pulse," Electron. Lett. **33**, 1080–1081 (1997).
11. M. Zoboli and P. Bassi, "High spatial resolution OTDR attenuation measurements by a correlation technique," Appl. Opt. **22**, 3680–3681 (1983).
12. P. Healey and P. Hensel, "Optical time-domain reflectometry by photoncounting," Electron. Lett. **16**, 631–633 (1980).
13. Q. Zhao, L. Xia, C. Wan, J. Hu, T. Jia, M. Gu, L. Zhang, L. Kang, J. Chen, X. Zhang, and P. Wu, "Long-haul and high-resolution optical time domain reflectometry using superconducting nanowire single-photon detectors," Sci. Rep. **5**, 10441 (2015).
14. W. Eickhoff and R. Ulrich, "Optical frequency domain reflectometry in single-mode fiber," Appl. Phys. Lett. **39**, 693–695 (1981).
15. D. Uttam and B. Culshaw, "Precision time domain reflectometry in optical fiber systems using a frequency modulated continuous wave ranging technique," J. Lightwave Technol. **3**, 971–977 (1985).
16. B. Soller, D. Gifford, M. Wolfe, and M. Froggatt, "High resolution optical frequency domain reflectometry for characterization of components and assemblies," Opt. Express **13**, 666–674 (2005).
17. S. Venkatesh and W. V. Sorin, "Phase noise considerations in coherent optical FMCW reflectometry," J. Lightwave Technol. **11**, 1694–1700 (1993).
18. F. Ito, X. Fan, and Y. Koshikiya, "Long-range coherent OFDR with light source phase noise compensation," J. Lightwave Technol. **30**, 1015–1024 (2012).
19. R. C. Youngquist, S. Carr, and D. E. N. Davies, "Optical coherence-domain reflectometry: a new optical evaluation technique," Opt. Lett. **12**, 158–160 (1987).

20. E. A. Swanson, D. Huang, M. R. Hee, J. G. Fujimoto, C. P. Lin, and C. A. Puliafito, "High-speed optical coherence domain reflectometry," Opt. Lett. **17**, 151–153 (1992).

21. K. Hotate, M. Enyama, S. Yamashita, and Y. Nasu "A multiplexing technique for fibre Bragg grating sensors with the same reflection wavelength by the synthesis of optical coherence function," Meas. Sci. Technol. **15**, 148–153 (2004).

22. Z. He, Zuyuan, T. Tomizawa, and K. Hotate. "High-speed high-reflectance-resolution reflectometry by synthesis of optical coherence function," IEICE Electron. Express **3**, 122–128 (2006).

23. Z. He, M. Konishi, and K. Hotate, "A high-speed sinusoidally frequency-modulated optical reflectometry with continuous modulation-frequency sweeping," Proc. SPIE **7004**, 70044L (2008).

24. K. Hotate and O. Kamatani, "Reflectometry by means of optical-coherence modulation," Electron. Lett. **25**, 1503–1505 (1989).

25. Z. He and K. Hotate, "Distributed fiber-optic stress-location measurement by arbitrary shaping of optical coherence function," J. Lightwave Technol. **20**, 1715–1723 (2002).

26. K. Hotate, "Application of synthesized coherence function to distributed optical sensing," Meas. Sci. Technol. **13**, 1746–1755 (2002).

27. Z. He, H. Takahashi, and K. Hotate, "Optical coherence-domain reflectometry by use of optical frequency comb," Conference on Lasers and Electro-Optics 2010 (CLEO2010), CFH4.

28. H. Takahashi, Z. He, and K. Hotate, "Optical coherence domain reflectometry by use of optical frequency comb with arbitrary-waveform phase modulation," 36th European Conference and Exhibition on Optical Communication (ECOC2010), Tu.3.F.4.

29. K. Hotate and K. Kajiwara, "Proposal and experimental verification of Bragg wavelength distribution measurement within a long-length FBG by synthesis of optical coherence function," Opt. Express **16**, 7881–7887 (2008).

30. D. Iida and F. Ito, "Detection sensitivity of Brillouin scattering near Fresnel reflection in BOTDR measurement," J. Lightwave Technol. **26**, 417–424 (2008).